# Investigating the effectiveness of multimodal data in forecasting SARS-COV-2 case surges

Palur Venkata Raghuvamsi[1#], Siyuan Brandon Loh[4], Prasanta Bhattacharya[4], Joses Ho[1], Raphael Lee Tze Chuen[1], Alvin X. Han[5], Sebastian Maurer-Stroh[1,2,3]

[1]Bioinformatics Institute (BII), Agency for Science, Technology and Research (A*STAR), 30 Biopolis Street, #07-01 Matrix Building, Singapore 138671, Republic of Singapore
[2]Department of Biological Sciences, Faculty of Science, National University of Singapore, 16 Science Drive 4, Block S3 #05-01, Singapore 117558, Republic of Singapore
[3]Human Potential Translational Research Program, Yong Loo Lin School of Medicine, National University of Singapore, 10 Medical Drive, Singapore 117597, Republic of Singapore
[4]Institute of High Performance Computing (IHPC), Agency for Science, Technology and Research (A*STAR), 1 Fusionopolis Way, #16-16 Connexis, Singapore 138632, Republic of Singapore.
[5]Department of Medical Microbiology and Infection Prevention, Amsterdam University Medical Center, University of Amsterdam, Amsterdam, The Netherlands
# - Corresponding author

## Abstract

The COVID-19 pandemic response relied heavily on statistical and machine learning models to predict key outcomes such as case prevalence and fatality rates. These predictions were instrumental in enabling timely public health interventions that helped break transmission cycles. While most existing models are grounded in traditional epidemiological data, the potential of alternative datasets, such as those derived from genomic information and human behavior, remains underexplored. In the current study, we investigated the usefulness of diverse modalities of feature sets in predicting case surges. Our results highlight the relative effectiveness of biological (e.g., mutations), public health (e.g., case counts, policy interventions) and human behavioral features (e.g., mobility and social media conversations) in predicting country-level case surges. Importantly, we uncover considerable heterogeneity in predictive performance across countries and feature modalities, suggesting that surge prediction models may need to be tailored to specific national contexts and pandemic phases. Overall, our work highlights the value of integrating alternative data sources into existing disease surveillance frameworks to enhance the prediction of pandemic dynamics.

## Introduction

The emergence and global spread of SARS-CoV-2 have caused profound disruptions to public health, economies, and daily life across all nations [1-3]. According to the WHO estimates, the pandemic resulted in approximately 775 million confirmed infections and nearly 7 million deaths, attributed to wild type and novel variants of the virus [4]. The mutations accumulated on the spike protein enabled the virus to escape host immune recognition, increasing transmissibility, and resulting in recurring infection surges. In response, governments worldwide swiftly introduced policies targeted at reducing transmission, such as mandatory mask usage, social distancing, closure of public spaces, and guidelines advising self-regulation of activities etc. [5]. The induced behavioral changes as a result of these policies make it challenging to empirically disentangle the effects of biological factors (e.g., viral fitness attained by natural evolution) and behavioral factors (e.g., human mobility and transmission) on the resulting surge in infections. Prior efforts to forecast pandemic outcomes (e.g., case counts) to plan for appropriate responses have been critical for healthcare professionals and regulators [6-9]. These models were either based on the efficacy of mutations that enhanced viral fitness i.e., antibody escape or tight binding human ACE2 receptor, or on epidemiological factors like case counts, hospitalization rates, test positivity rates etc. [10, 11].

Studies using these modeling approaches have reported high predictive accuracy, leading to their adoption in pandemic response efforts. For instance, the SEIR (Susceptible–Exposed–Infectious–Recovered) and its variants were widely adopted as primary epidemiological models to forecast disease trends during the pandemic [12, 13]. These models have been successful in predicting relevant public health outcomes like reproduction rate of the disease, i.e., the expected number of secondary infections produced by a single (typical) case in a fully susceptible population. However, compartmental models such as the SEIR are often limited by the inherent assumption of population homogeneity, and the need for accurate characterization of epidemiological features like transmissibility or infectiousness of the virus etc. [8]. In other studies, autoregressive time series models have been employed to forecast case counts and surges using historical case data [14, 15]. Recently, time series-based machine learning (ML) models have been successfully used to forecast pandemic trends using different modalities of data. ML models, in particular, have been instrumental in predicting both clinical and epidemiological outcomes such as mortality, hospitalization rates, surge in cases etc. from large secondary datasets [8, 16, 17].

Such ML-based models are particularly valuable during the later stages of a pandemic, as well as retrospectively, when the quality of epidemiological data, such as case and death counts, is generally more reliable. However, forecasting outcomes remains particularly challenging during phases when public health data is sparse, such as (a) in the early stages of a pandemic, (b) in regions with limited testing or inconsistent data collection, or (c) during dormant phases marked by low case counts due to stringent lockdown measures. In such scenarios, relying on a multimodal data strategy which combines information from multiple biological and behavioral factors that directly or indirectly proxy viral transmissibility can be useful for guiding policy response. However, the effectiveness of incorporating diverse data modalities into predictive public health models remains understudied [18].

As such, the current study seeks to answer two primary questions: (1) Can multimodal models, utilizing a combination of biological and behavioural modalities, outperform unimodal models in predicting case surges? and (2) Are there heterogeneities in multimodal model performance depending on the specific country or modality being studied? To answer these, we curated a novel dataset encompassing both biological and human behavioral features, including: I) SARS-CoV-2 mutation data from the COVID-19 GISAID database, (II) COVID-19 case and death counts, (III) Policy responses, measured by the Stringency Index (SI), (IV) Offline behavior, represented by population mobility data from Google's COVID-19 community mobility reports, and (V) Online social media behavior, captured through emotion features extracted from a global Twitter dataset.

Our findings reveal substantial heterogeneity in model performance based on both country and data modality. Specifically, we find that incorporating bimodal feature combinations often improves predictive accuracy compared to unimodal models, but this improvement is not consistent across all countries. Our results also highlight the effectiveness of alternative modalities of data, such as social media-based emotions and mobility, which are *indirectly* related to case surges unlike case or death counts which offer a more direct indication of disease transmission. In data-scarce public health contexts, such as the initial phases of a disease outbreak, ML models trained on diverse and alternative modalities of data can help policy makers initiate timely and informed responses.

## Methods

### Dataset collection and preparation

The Our World in Data (OWID) COVID-19 repository [19] provides multiple modalities of variables including confirmed new cases, deaths, ICU patients, vaccinations, stringency index etc. We obtained complete daily data from OWID covering the full period of the SARS-CoV-2 outbreak, and computed weekly averages for each variable from September 2019 to November 2022. For this study, we focused on data from seven countries: United States of America, United Kingdom, India, Singapore, Malaysia, and Germany. We explicitly chose this temporal window to obtain a clear differentiation of pre- and post- vaccination periods for the countries studied. Tables 1 and 2 summarize all features used to predict case surges, categorized into five data modalities, which were analyzed both individually and in combination. To enable further analysis, a weekly surge ratio was calculated using Eq. 1 below [20, 21].

$$Surge\ Ratio_t = \frac{Case\ count_{t-1}}{Case\ count_{t-2}} \quad (1)$$

The surge ratios were subsequently discretized into surge labels, hereafter referred to as $SL_0$, using threshold values as shown in Eq. 2.

$$SL_0 = \begin{cases} a, & 0 < Surge\ ratio \leq 0.5 \\ b, & 0.5 < Surge\ ratio \leq 1.0 \\ c, & 1.0 < Surge\ ratio \leq 1.5 \\ d, & 1.5 < Surge\ ratio \leq 2.0 \\ e, & Surge\ ratio > 2.0 \end{cases} \quad (2)$$

These labels represent the intensity of surge in case counts resulting from multiple factors such as the emergence of a new variant, introduction of or changes to COVID-19 policies, shifts in population mobility patterns etc. Next, we illustrate the key modalities of data that we use for this study.

***Case counts*** Daily confirmed SARS-CoV-2 case and death counts were obtained from the Our World in Data (OWID) repository [19].

***Policy response*** Government responses to pandemic in terms of restricted mobility, lockdowns, health care policies etc. were captured using the Stringency Index (SI) [22], a composite metric reflecting the overall strictness of a country's COVID-19 policies. The SI data, provided by the Oxford COVID-19 Government Response Tracker and available through the OWID repository, were collected at a daily basis and aggregated into weekly averages for each country.

***Genomic mutational data*** The GISAID database is the primary data repository for collating SARS-CoV-2 genome sequences [23-25]. We utilized an in-house tool to translate all the genomic sequences into protein [26]. Further we calculated various parameters such as weekly mutational counts in the whole proteome, Spike protein alone, and RBD alone with respect to wild type. A total of 10 features were generated from the GISAID sequence dataset [26].

***Social media-based emotions*** Twitter data was obtained from an expanded version of the dataset used in [27]. This global dataset includes COVID-19 related public tweets by users from a host of countries and is labelled on a number of sentiment and emotion-related attributes. We included the intensity of fear, anger, sadness and happiness emotions for the current analysis.

***COVID-19 community mobility reports*** Aggregated mobility change data from the publicly released Google community mobility reports was used to estimate the public mobility trends [28] across location types (e.g., residential vs. workplace). A total of 6 features covering the various location types such as public commute to work, grocery, parks etc. were included for the current analysis to capture shifts in offline public behavior.

## Model development

For the classification task, we followed a sliding-window approach where aggregated multimodal data from prior weeks, specifically week t-1 ($t_{-1}$), t-2 ($t_{-2}$), or both, were used to predict $SL_0$ for the target week t. The feature modalities were provided individually (i.e., for unimodal models) or in combination (e.g., in bimodal models) to train a set of supervised ML models for $SL_0$ prediction. In addition to classification, we evaluated the effectiveness of the feature modalities in a *regression* task aimed at predicting the absolute value of the surge ratio for week t, hereafter referred to as $SR_0$. For both tasks, all input features were standardised. We used an 80-20 train-test split, where data from 28th September 2019 to 16th May 2022 was used for training, while data from 16th May 2022 to 17th October 2022 was held back for testing. Missing values in both feature sets were imputed using forward fill (ffill) and backward fill (bfill) methods.

We implemented 4 machine learning models for the regression and classification tasks: Extreme Gradient Boosting (XGBoost), Random Forest (RF), Support Vector Machine (SVM) and a two-layer Long Short-Term Memory (LSTM) Network with 50 units per layer. All models were implemented using the scikit-learn python package or PyTorch (for LSTM) [31].

## Model evaluation

The raw accuracy (%) and F1-score metrics were used to evaluate model performance on the classification task, while Mean Absolute Error (MAE) was used to evaluate model performance on the regression task.

To assess the relative improvement of bimodal features in the prediction task, we used the MM1 metric [32], commonly used in the affective computing field to quantify the performance gains of multimodal emotion recognition models over unimodal baselines. Specifically, we use a country-level (or) country/modality-level MM1 metric to quantify the improvement from using multimodal models (over unimodal models). This metric measures the difference between the best (or mean) performing multimodal model and the best (or mean) performing unimodal model, as specified in Eq. 3-6 below. $MM1_C$ and $MM1_R$ scores represent classification- or regression-specific MM1 scores respectively.

$$MM1_{C,Best,\,country,modality} = [\![\min\{MAE]\!]_{uni-modal,\,country,modality}\} - min\{MAE_{multi-modal,country,modality}\} \qquad (3)$$

$$MM1_{C,mean,\,country,modality} = [\![\text{mean}\{MAE]\!]_{uni-modal,\,country,modality}\} - mean\{MAE_{multi-modal,country,modality}\} \qquad (4)$$

$$MM1_{R,best,\,country,modality} = [\![\max\{Accuracy]\!]_{uni-modal,\,country,modality}\} - max\{Accuracy_{multi-modal,country,modality}\} \qquad (5)$$

$$MM1_{R,mean,\,country,modality} = [\![\text{mean}\{Accuracy]\!]_{uni-modal,\,country,modality}\} - mean\{Accuracy_{multi-modal,country,modality}\} \qquad (6)$$

# Results

## Predicting COVID-19 case surges

As described in the previous section, we curated a dataset integrating multiple data modalities spanning genomic (i.e., mutation counts), public health (i.e., case and death counts), mobility (i.e., location-based mobility change %), and social media (i.e., Twitter emotion intensities) data from various sources, and used in-house tools to generate the relevant features at a country level (See Tables 1 and 2). These features serve as direct or indirect indicators of viral infection and transmission, and can be suitably utilized to forecast the intensity of upcoming case surges (See Table 2, Figure 1 and Figure S1-S4). Recent studies have highlighted trends in these features during the pandemic [14, 24]. In the current study, we computed weekly averages for each feature as well as for the target variables (i.e., Surge Ratio ($SR_0$) and Surge Label ($SL_0$)) of the prediction tasks. $SR_0$ was calculated as the ratio of the case counts of preceding 1st week ($t_{-1}$) and 2nd week ($t_{-2}$) (See Figure 2 and Figure S5). An $SR_0$ value of lower or equal to 1 implies a stable or declining case count, while values greater than 2 are indicative of a notable surge in cases. As expected, substantial increases in $SR_0$ values in our data correspond to key pandemic events such as the introduction of novel variants (e.g., Delta and Omicron) into the population, or the enforcement of public health interventions such as lockdowns (See Figure S5). Further, we investigated three different approaches for discretizing $SR_0$ into categorical Surge Labels ($SL_0$):

**Two class scheme:** *No surge*: $0 < SR_0 < 1$ | *Surge*: $SR_0 > 1$

**Three class scheme:** *No surge*: $0 < SR_0 < 1$ | *Moderate surge*: $1 < SR_0 < 2$ | *High surge*: $SR_0 > 2$

**Five class scheme**: *No surge*: $0 < SR_0 < 0.5$ | *Low surge*: $0.5 < SR_0 < 1$ | *Moderate surge*: $1 < SR_0 < 1.5$ | *Surge*: $1.5 < SR_0 < 2$ | *High surge*: $SR_0 > 2$

From a public health planning perspective, we contend that similar $SR_0$ values (e.g., 1.5 vs. 1.7) are likely to require comparable levels of resource allocation. Accordingly, we trained ML models to predict $SL_0$ i.e., the discretized surge labels derived from $SR_0$, using the classification scheme described above. These labels allow for a more actionable interpretation of case surges across different $SR_0$ ranges.

To evaluate the relative efficacy of different feature modalities in classifying surge labels, we trained the 4 supervised machine learning models as detailed in the previous section. Each model was trained using all feature modalities, lagged by both 1st ($t_{-1}$) and 2nd ($t_{-2}$) week, to predict the current week's ($t_0$) surge labels ($SL_0$). Our results revealed considerable heterogeneity in test performance based on model type and specification (See Figure 3A). However, and across countries, the XGBoost model performed consistently better than other models, and hence we used this model for all subsequent analyses in this study, as detailed next.

### Performance heterogeneities in predicting $SR_0$ and $SL_0$

We trained an XGBoost model with above-mentioned feature modalities, added individually, to predict the current week's ($t_0$) surge labels using features from the preceding week (i.e., $t_{-1}$) or two weeks prior (i.e., $t_{-2}$) or a combination of both weeks (i.e., $t_{-1} + t_{-2}$). We used raw accuracy and F1 scores to compare efficacy of forecasting using the different feature modalities (See Figures 3B-3F and Figure S6). Interestingly, comparison of raw accuracy and F1 scores revealed significant country-wise heterogeneity in predicting $SL_0$ with respect to feature modality and choice of lag period (See Figure S6). For instance, in the case of the UK, none of the feature modalities were sufficiently predictive except for the mutation feature set (See Figures 3B-3F), highlighting the dominant role of genomic data in classifying surge labels for that country. Similarly, the mutation modality yielded higher accuracy in predicting $SL_0$ for Singapore compared to other feature modalities. The models trained on features from other countries performed with reasonable accuracy, with mutation and mobility feature sets from $t_{-1} + t_{-2}$ lag yielding highest accuracy for Malaysia and USA, respectively. Notably, the policy modality, measured by the stringency index, showed no variability in predictive value across different lag periods for any country (Figure 3E). We conjecture that this might be due to inherent delays in behavioural change resulting from government directed policies or initiatives that often take time to onset, and can last longer than the 1-2 week period that we study here. As such, models trained on data lagged by only 1-2 weeks might not effectively capture the effect of such policy actions (e.g., mobility restrictions or school closures). Nonetheless, model performance varied in a country-wise manner highlighting the differing intensities of policy responses to the pandemic.

Models trained using feature modalities from the USA and Malaysia achieved accuracies of at least 0.5, with the mobility feature set from $t_{-1}$ + $t_{-2}$ lag weeks yielding the highest accuracy. This emphasizes the utility of including a wider range of modalities in predicting the case surge (See Figures 3B-3F). The country-wise variations in model performance hint at the variability in general responses to the pandemic, and the importance of adopting a multimodal and country-specific approach to predicting pandemic outcomes.

To further assess the performance of XGBoost models, we employed probability-based gradient boosting methods to derive the probability of assigning each $SL_0$. The assignment of a probability to each label allowed us to infer the extent of deviation from the true label (See Figure 4 and Figures S8-S21). Figure 4 shows an illustration where the probability of classification of each $SL_0$ by the XGBoost model trained on tweet emotion data is presented, along with its accuracy and F1 scores. Moreover, to effectively address class imbalance in predicting the surge label, we generated model predictions using two alternative surge labeling schemes as discussed earlier i.e., three class vs. two class scheme. Consistent with the performance of models with finer surge labels i.e. five class scheme, models trained under the three or two class schemes also show country-wise variations in performance. Interestingly, the prediction of finer labels showed better performance than coarse surge labels based on F1 scores (See Figure S7), suggesting that our model is capable of distinguishing finer variations in surge intensity.

Finally, as a robustness check of our outcome measure, we trained models to directly predict the continuous surge ratio ($SR_0$) instead of the discretized surge label ($SL_0$), and used Mean Absolute Error (MAE) to compare the model performance. Models trained with all 5 modalities of features from both the lag weeks show country-specific performance that are largely consistent with the $SL_0$ classification task (See Figure 5). Further, the models trained on individual feature modalities, using lag weeks $t_{-1}$ or $t_{-2}$ or $t_{-1}$ + $t_{-2}$ showed satisfactory performance with MAE values ranging between 0.15-0.4 depending on the country, except for models trained on data from India. Unlike the $SL_0$ classification task, the $SR_0$ predictions showed less variability in MAE between lag weeks. Furthermore, we illustrate the relative importance of features in predicting the target variable in both classification and regression tasks (See Figures S22-S29).

**Phase-specific model performance**

The predictive performance of different data modalities likely varied substantially over the course of the pandemic due to factors such as the evolution of viral strains, differences in country-level responses in the form of vaccination campaigns, testing rates, travel restrictions etc., as well as unobserved variances in data quality and availability.

For instance, in the initial stages of the pandemic, mutation data was sparse due to delays in response, dissemination of public information, and resource allocation for sequencing viral samples from patients. Similarly, the availability and quality of COVID-19 related social media data were likely lower in the early stages as compared to the later stages of the pandemic. To account for such variations, we tested model performance based on each feature category across four distinct six-month phases between 2020 and 2022. For this exercise, we used data from Singapore and UK, as these two countries have sufficiently differentiated characteristics in terms of their population profiles and pandemic responses.

We specifically evaluated changes in $SL_0$ classification performance across the four six-month pandemic phases (See Figure 7). For Singapore, our analyses shows that case and death count features performed the best, with performance improving markedly in later phases compared to the initial phases (See Figure 7A). Conversely, the predictive performance of models trained on mutation features declined over time, from 0.66 in Phase 1 to 0.50 in Phase 4 (See Figure 7A(i)), displaying temporal dynamics akin to that observed for case and death count features. In the case of the UK, model performance was highest during the first phase of the pandemic, with the performance gradually reducing to varying degrees depending on the modality, except for stringency index (See Figure 7B). Interestingly, our results show that models trained on the social media features exhibited the least reduction in accuracy from the initial to the final phase of the pandemic i.e., 0.66 to 0.5 (See Figure 7B(i-iii)). The reduction in model performance for the classification task during the final stage of the pandemic (i.e., Q4) may be attributed to various unobserved factors, such as increased vaccination, immunity developed from previous exposures etc. Nonetheless, our overall analysis shows the utility of leveraging multimodal datasets in capturing the effects of the pandemic, as well as the impacts of country-specific responses at various stages of the pandemic.

**Heterogeneities in multimodal model performance**

As demonstrated in earlier sections, predictions using unimodal feature modality sets showed varying levels of performance depending on the country and the specific feature modality being employed. Drawing on this, we next evaluated model performance in forecasting $SL_0$ using bimodal features. Specifically, we hypothesized that combining features across modalities will likely enhance model performance in predicting $SL_0$, but with notable heterogeneities in improvement across countries, choice of temporal lags, and the type of modalities used. We constructed a total of 10 bimodal feature sets from three lag week periods for this task (See Table3). We then calculated the MM1 score to quantify the additive effect of using bimodal features over unimodal features (See Figure 6 and Figure S30). Next, the models trained on the bimodal feature combinations were used for the regression task of predicting $SR_0$. As expected, both regression and classification tasks showed country-wise and lag week specific heterogeneities in accuracy and MAE respectively (See Figure 6 and Figure S31). We calculated both $MM1_C$ and $MM1_R$ metrics using raw accuracy scores for the classification task, and MAE for the regression task (See Figure 8). As previously described, the MM1 metric measures the difference in performance between the best (or mean) performing multimodal model and the best (or mean) performing unimodal model.

Interestingly, for the classification task, countries that performed poorly when trained on unimodal features showed improvement when trained using bimodal features. For instance, in the case of UK and Brazil, models trained on bimodal features predicted surge labels more effectively, as shown in Figures 8A-8B. Notably, models trained on data from the USA and UK showed opposing trends, with bimodal features improving task performance for the UK but reducing task performance for the USA. Across both countries, we noticed that models trained on features lagged by both $t_{-1}$+ $t_{-2}$ periods outperformed those using features from either $t_{-1}$ or $t_{-2}$ lag periods, in predicting $SL_0$. In case of the regression task, and based on the $MM1_R$ scores, we observed that models trained on unimodal features performed better than those trained on bimodal features. However, it is worth noting that a $MM1_C$ sore difference of 0.1 corresponds to 10% change in accuracy whereas the same magnitude of difference in a regression task i.e. $MM1_R$ is defined as the difference in absolute $SR_0$ value measured in MAE. Therefore, in regression tasks, unimodal feature modalities performed marginally well compared to bimodal feature modalities.

## Discussion

Timely forecasting of pandemic outcomes is critical for launching early interventions and informing public health policy. The time to mount an effective response should ideally be weeks ahead of a predicted case surge. In this regard, developing predictive models to forecast key public health outcomes (e.g., cases, deaths, hospitalizations etc.) is crucial, and multiple studies have shown the utility of such models. However, since the availability and quality of conventional public health data in the early stages of the pandemic can be uncertain, the use of alternative datasets offer a possible solution.

Previous studies have used machine learning or statistical models leveraging individual data modalities, such as mobility, policy, or social media indicators, to forecast case outcomes. Our current study builds on this past work in two key ways. First, we adopt a multimodal modeling approach, combining data from diverse modalities such as genomics, policy, mobility, and social media, to improve the prediction of case surges. Second, we evaluate the relative performance of unimodal and bimodal feature modalities in predicting case surges, to help inform policy decisions on their relative effectiveness. Our results highlight significant country-level heterogeneity in pandemic trends and corresponding performance of ML models in predicting case surges. Further, by leveraging an explainable model we identified important features within each modality, used individually or in combination, in terms of their contributions to predicting case surges.

Past studies have shown a direct link between some of the policy and behavioral factors that we study in this paper, and pandemic related outcomes [33, 34]. For instance, lockdowns that severely restrict population mobility are a particularly stringent form of policy response and have been shown to be effective in breaking the transmission chain, reducing the number of cases [35, 36]. Our use of the stringency index, as a composite measure of such policy responses, helps capture the aggregate severity of these decisions. Further, inherent evolutionary dynamics of the SARS-CoV-2 through accumulation of mutations due to selection pressures or during the replication cycle positively affect viral fitness. During the pandemic, periodic introductions of novel SARS-CoV-2 variants have often resulted in drastic rise in case counts, as observed in the case of alpha, delta and omicron variants. Identifying the emergence of a novel variant in near real-time is a non-trivial task and depends on both the early detection of a drastic surge in case count, coupled with experimental genomic validation to pinpoint the variant-defining mutation. To account for the emergence of novel variants via introduction of point mutation in the viral genome, we computed multiple features related to point mutations reported in GIASID database onSARS-CoV-2. These features help proxy for the biological infectiousness of the variant and complement the other behavioral modalities of data that we use in our model.

Another important contribution of the current study is our incorporation of both offline and online behaviors in the form of mobility change and social media emotions, respectively. These behavioral indicators help capture the risk of disease transmission in the population. For instance, increased mobility creates higher opportunities for individuals to co-locate and infect each other. Similarly, global social media emotions might reflect a combination of users' affective responses to the onset and growth of the pandemic, as well as their responses to the policy measures implemented. Interestingly, we found that for certain countries, models trained on online and indirect measures of pandemic outcomes (e.g., social media emotions) outperformed models trained on direct public health measures. However, and as an important caveat, we note that while the integration of multiple modalities of public health and alternative datasets can contribute to disease surveillance, it is important to be mindful of the inherent performance variabilities that exist across countries and data modalities. We recommend that country-specific ML models trained on genomic and human behavioral data can be used together with, or as complements to existing epidemiological models to better forecast pandemic outcomes.

## Supplementary Information

Supplementary information is available at Zenodo: "https://doi.org/10.5281/zenodo.15469321"

## Acknowledgments

This work was supported by fundings from A*STAR Epidemic Preparedness Horizontal Coordinating Office (EP HTCO) FY22_CF_HTCO_SEED_EP_BII_C221418002.The authors are grateful to Dr. Yang Yinping for her helpful suggestions on this paper, and to Dr. Raj Kumar Gupta for his assistance with curating the social media datasets. RV would like to acknowledge Bioinformatics Institute (A*STAR) core funds.

## Conflict of interest

Authors claim no conflict of interest

Table 1: Feature source and summary

| Feature type | Source | Number of features |
|---|---|---|
| **Case and deaths** | Our World in Data (OWID) | 12 |
| **Mobility** | Google Mobility | 6 |
| **Mutations** | EpiCOV/GISAID | 10 |
| **Policy stringency** | OWID | 1 |
| **Social media emotions** | Gupta et.al 2021 | 9 |
| **Total** | | 38 |

Table 2: Features in each modality

| Case counts | Mobility | Mutations | Policy | Social media |
|---|---|---|---|---|
| **1.** Total cases<br>**2.** New cases<br>**3.** New cases smoothed<br>**4.** Total deaths,<br>**5.** New deaths<br>**6.** New deaths smoothed<br>**7.** Total cases per million<br>**8.** New cases per million<br>**9.** New cases smoothed per million<br>**10.** Total deaths per million<br>**11.** New deaths per million<br>**12.** New deaths smoothed per million | 1. Mobility retail and recreation percent change from baseline<br>2. Mobility grocery and pharmacy percent change from baseline<br>3. Mobility parks percent change from baseline<br>4. Mobility transit stations percent change from baseline<br>5. Mobility workplaces percentage change from baseline<br>6. Mobility residential percent change from baseline | 1. All mutations count mean<br>2. Spike mutations count mean<br>3. Prop Spike mutations mean<br>4. Spike_RBD mutations count mean<br>5. Prop Spike_RBD mutations mean<br>6. All mutations count approximate median<br>7. Spike mutations count approximate median<br>8. Prop spike mutations approximate median<br>9. Spike_RBD mutations count approximate median<br>10. Prop Spike_RBD mutations approximate median | 1.Stringency Index | 1. Tweet-n (total number of tweets),<br>2. Tweet-prop fear<br>3. Tweet-prop anger<br>4. Tweet-prop sad<br>5. Tweet-prop happy<br>6. Tweet-mean fear<br>7. Tweet-mean anger<br>8. Tweet-mean sad<br>9. Tweet-mean happy |

Table 3: Bimodal features combinations used for surge ratio or surge label predictions

| Bimodal feature | Total number of features |
|---|---|
| Mobility + Social media | 15 |
| Mobility + Cases | 18 |
| Mobility + Policy | 7 |
| Mobility+ Mutations | 16 |
| Social media + Cases | 21 |
| Social media + Policy | 13 |
| Cases + Policy | 13 |
| Cases + Mutations | 22 |
| Policy+ Mutations | 11 |

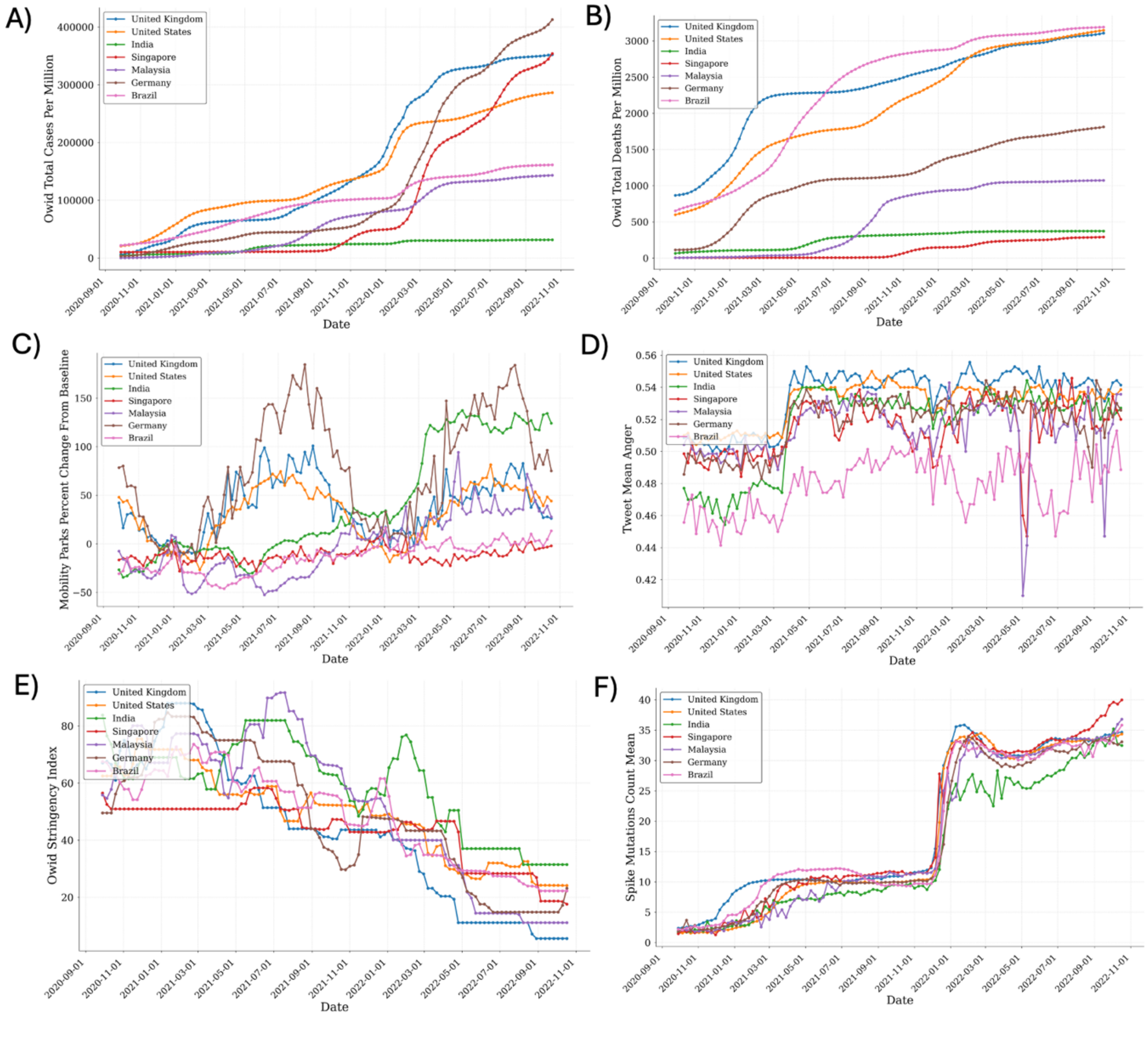

Figure 1: Weekly average of representative features from each modality for the pandemic period spanning September 2019 to December 2022

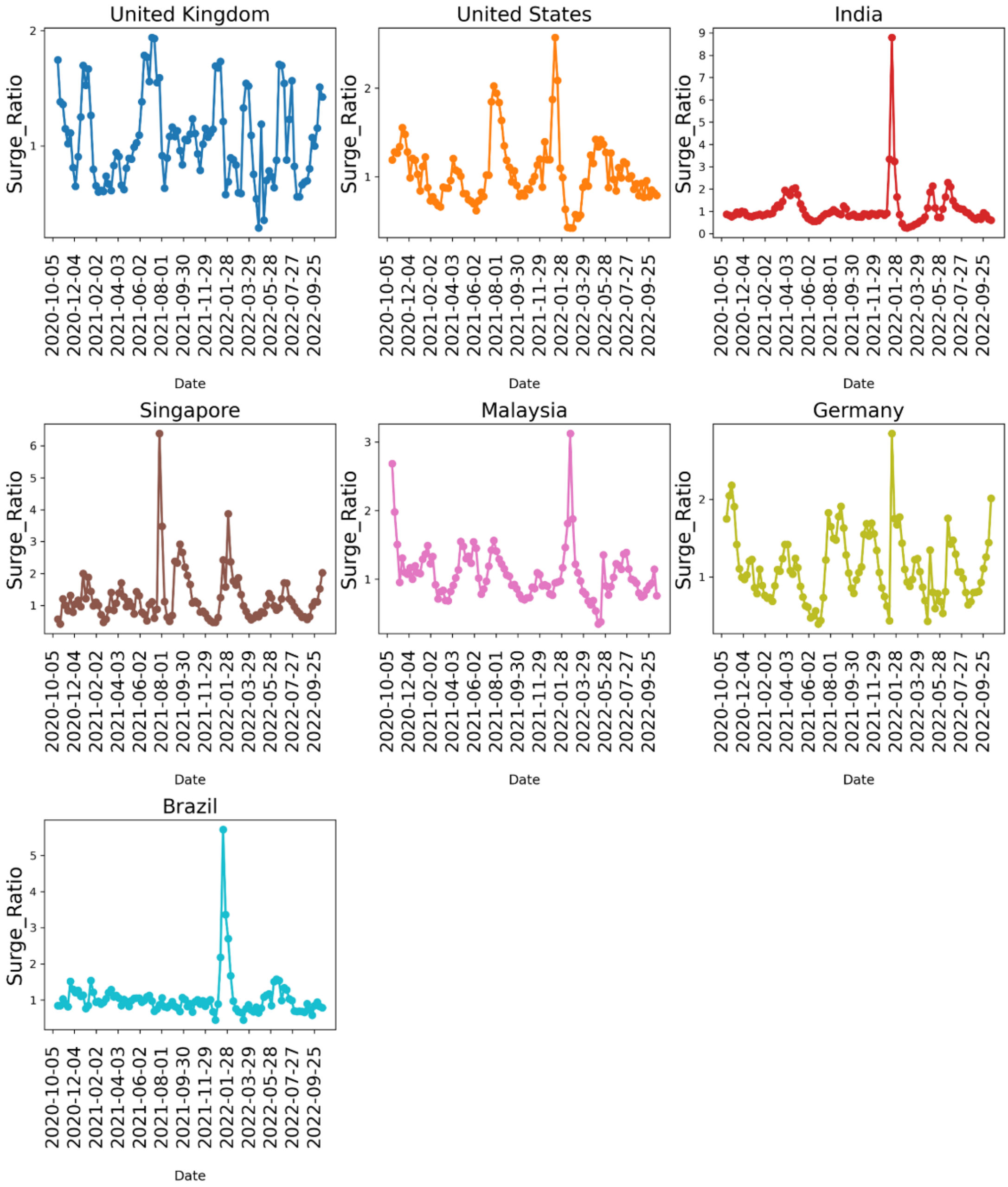

Figure 2: Weekly $SR_0$ value for each country for the pandemic period spanning September 2019 to December 2022

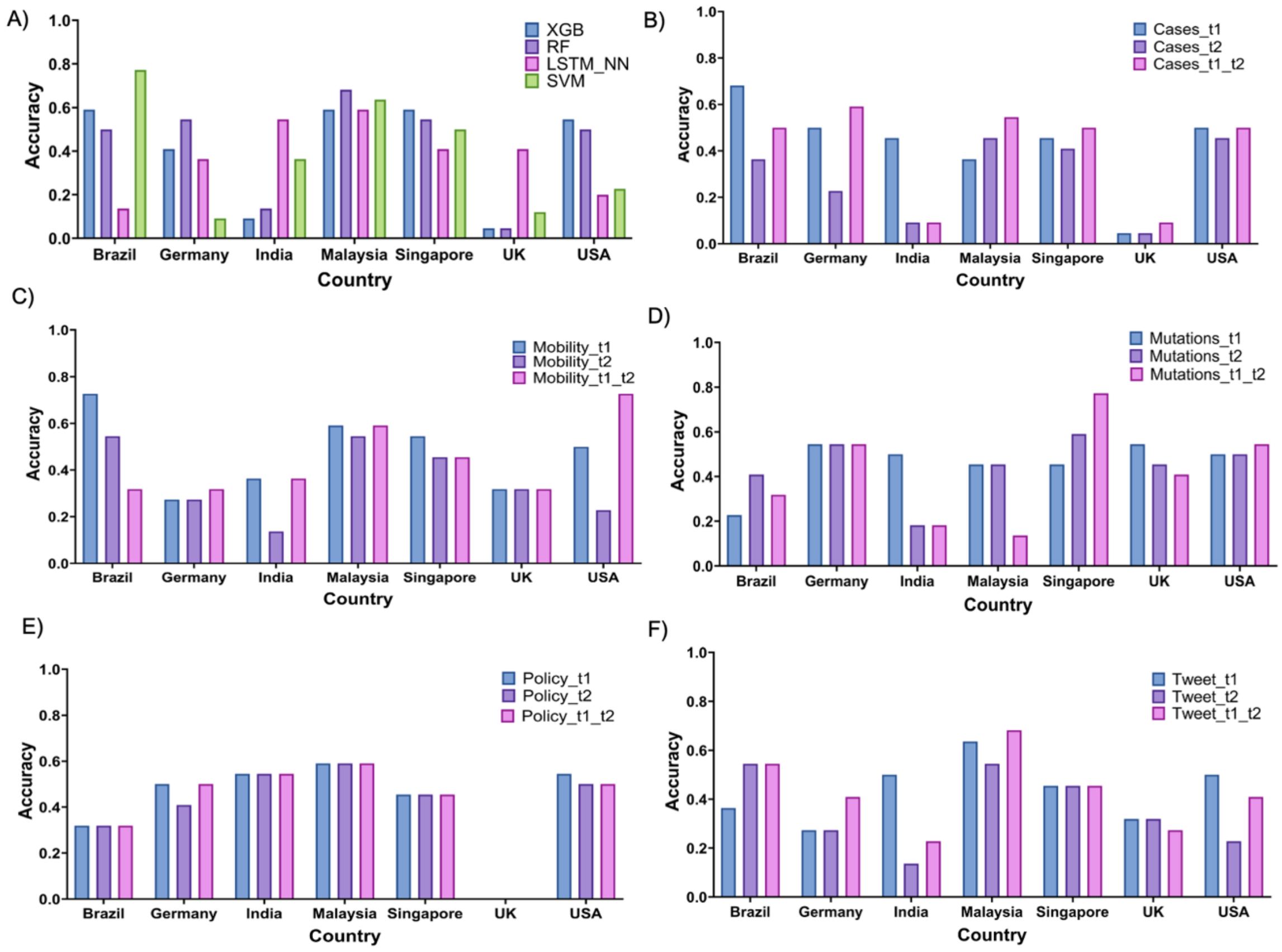

Figure 3: Comparing country-wise model performance in predicting $SL_0$ using different features from September 2019 to December 2022. A) Country-wise accuracy of models using all 5 modalities of features. B-F) Country-wise accuracy of models using each feature modality used as input, with different lags

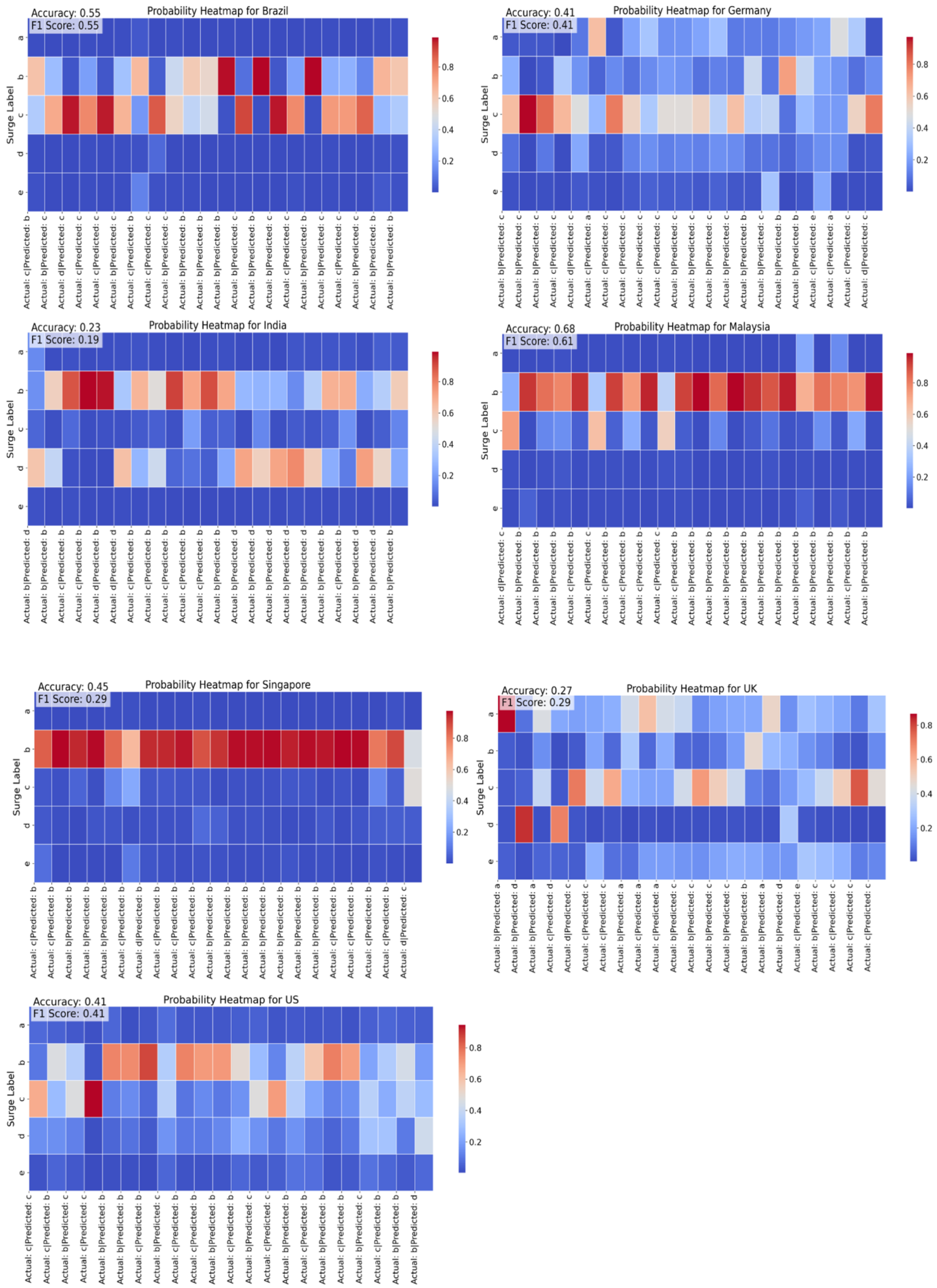

Figure 4: Label-wise probability assignment for the classification task trained on Twitter emotion feature. (Labels: a - no surge, b - low surge, c - moderate surge, d - surge and e – high surge)

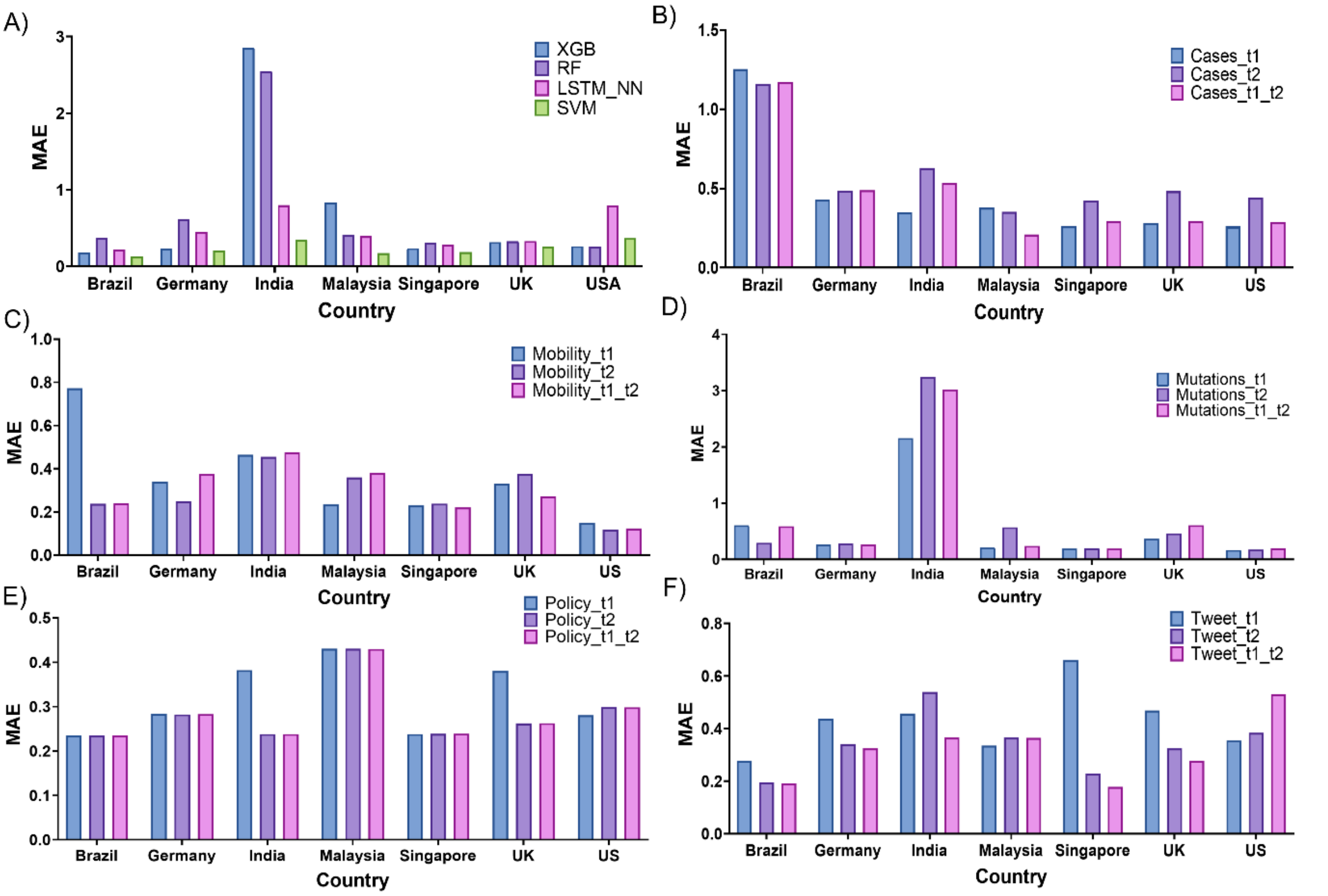

Figure 5: Comparing country-wise model performance in predicting $SR_0$ using different feature modalities A) Country-wise accuracy of models using all 5 modalities of features. B-F) Country-wise accuracy of models with each feature modality used with different time lags

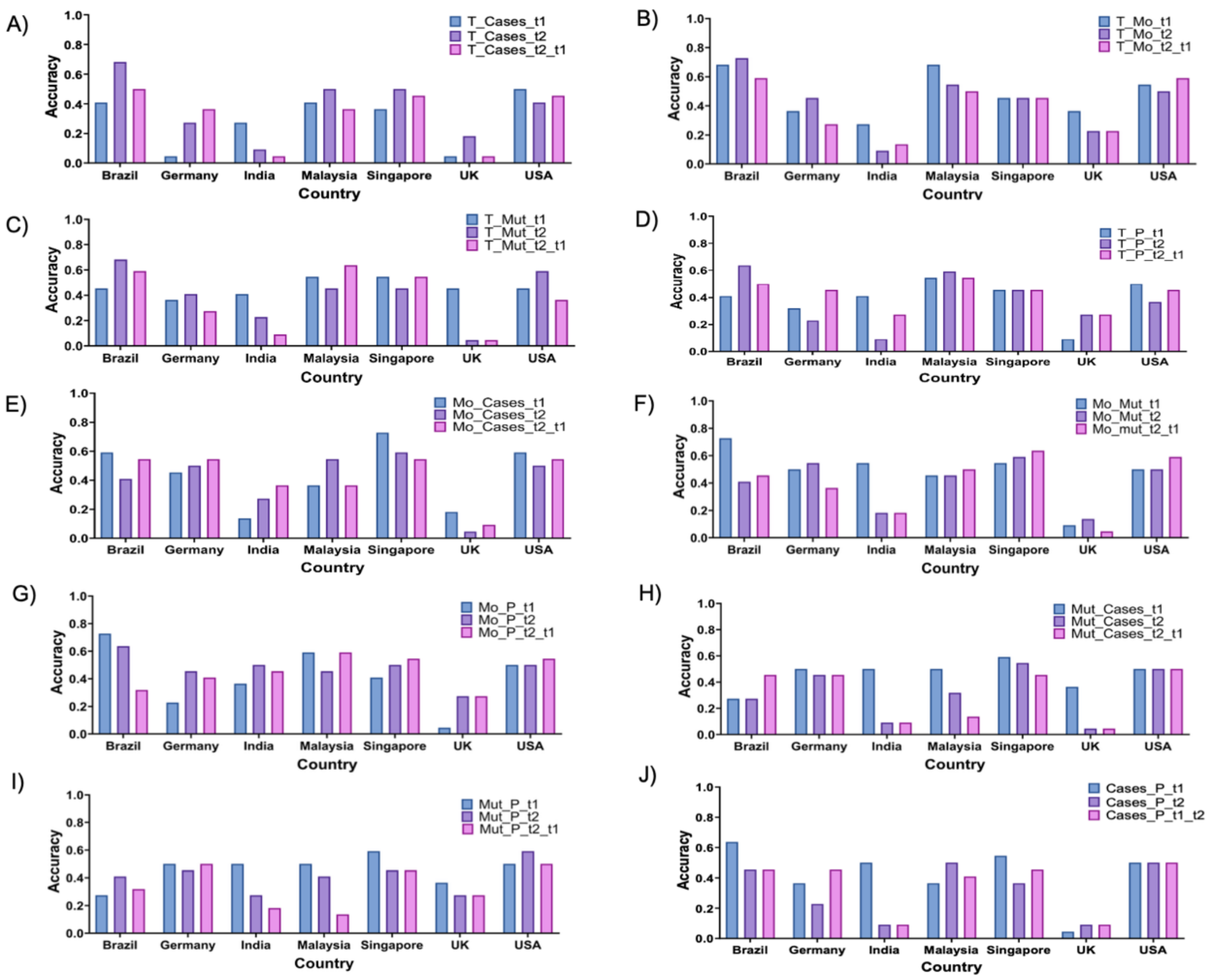

Figure 6: Comparing country-wise model performance in a 5-class classification of $SL_0$ using different combinations of bimodal features. A) Country-wise accuracy of models with all 5 modalities of features. B-F) Country-wise accuracy of models with each feature modality used with different time lags

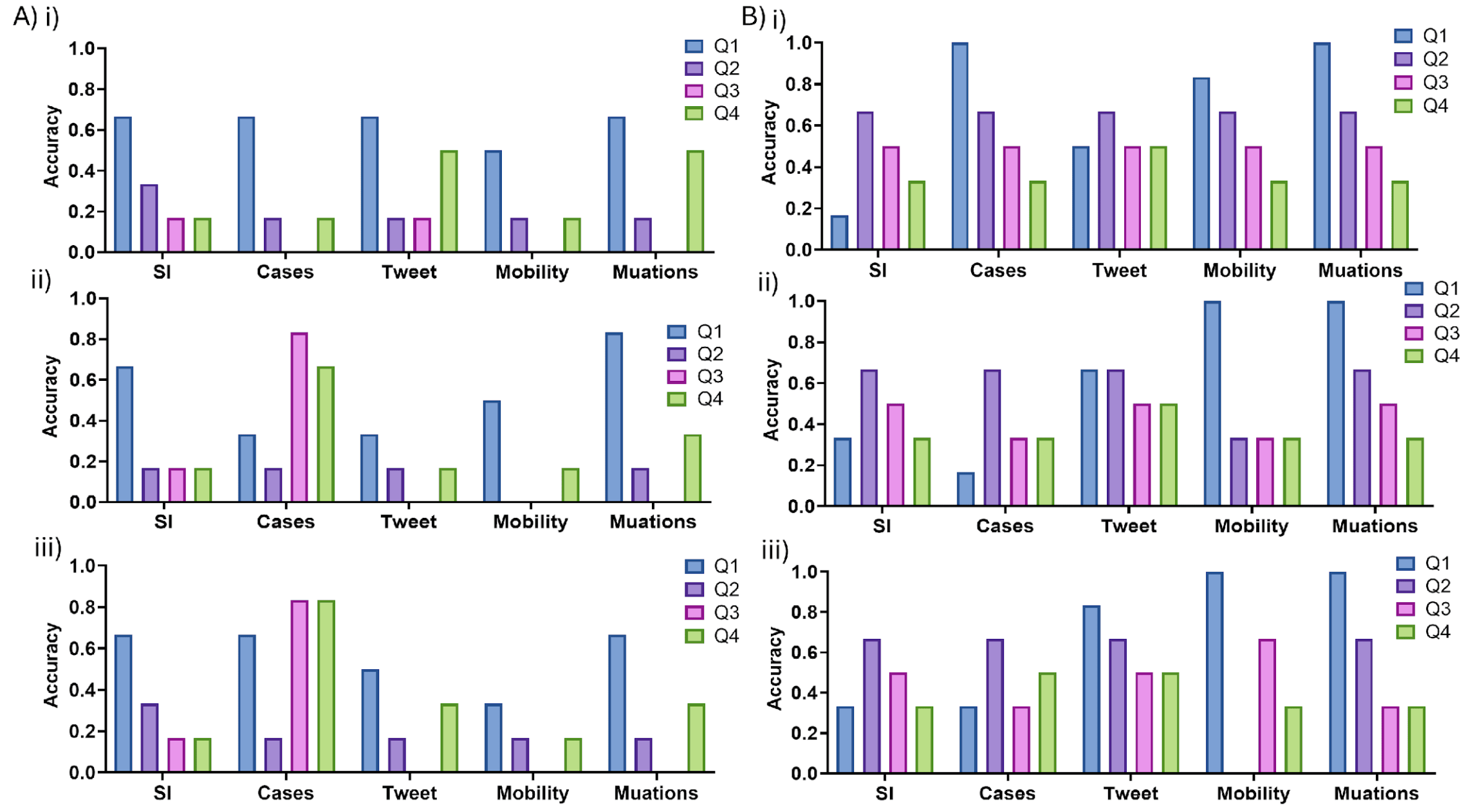

Figure 7: Comparing XGBoost model performance trained on each feature modality at different pandemic phases in a 5-class classification of $SL_0$. A) Modality-wise accuracy of models with 80-20 train-test split for each six-monthly subset of the Singapore dataset. i) Model trained on i) t1, ii) t2 and iii) t1+t2 lag weeks. B) Modality-wise accuracy of models with 80-20 train and test split for each six-monthly subset of the UK dataset. i) Model trained on i) $t_1$, ii) $t_2$ and iii) $t_1$+$t_2$ lag weeks

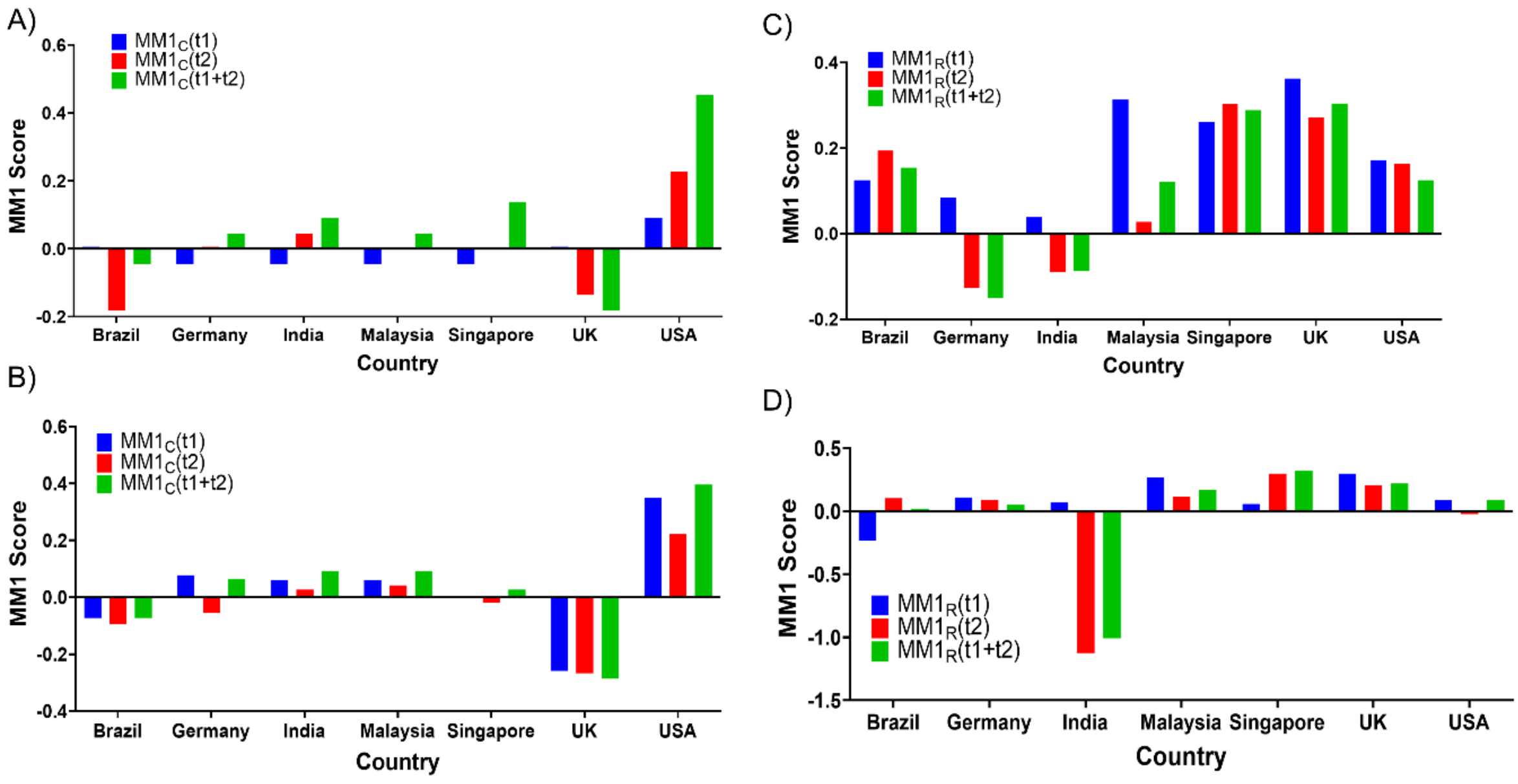

Figure 8: Comparing model performance with bimodal vs unimodal features in predicting $SL_0$ (5 class) and $SR_0$ using the MM1 score. A-B) Country-wise MM1 score using maximum accuracy (A) or mean accuracy (B) for model predicting $SL_0$. C-D) Country-wise MM1 score using minimum MAE (A) or mean MAE (B) for model predicting $SR_0$.